# Alternative solutions to caesium in negative-ion sources: a study of negative-ion surface production on diamond in H$_2$/D$_2$ plasmas

Gilles Cartry[1], Dmitry Kogut[1], Kostiantyn Achkasov[1], Jean-Marc Layet[1], Thomas Farley[2], Alix Gicquel[3], Jocelyn Achard[3], Ovidiu Brinza[3], Thomas Bieber[3], Hocine Khemliche[4], Philippe Roncin[4], Alain Simonin[5]

[1] Aix-Marseille Univ, CNRS, PIIM, UMR 7345, 13013 Marseille, France
[2] CCFE, Culham Science Centre, Abingdon, Oxon, OX14 3DB, UK and Department of Electrical Engineering and Electronics, University of Liverpool, Brownlow Hill, Liverpool, L69 3GJ, UK
[3] LSPM, CNRS-UPR 3407 Université Paris 13, 99 Avenue J. B. Clément, F-93430 Villetaneuse
[4] Institut des Sciences Moléculaires d'Orsay (ISMO), CNRS, Univ. Paris-Sud, Université Paris-Saclay, F-91405 Orsay, France
[5] CEA, IRFM, F-13108 Saint-Paul-lez-Durance, France
gilles.cartry@univ-amu.fr

## Abstract

This paper deals with a study of H-/D- negative ion surface production on diamond in low pressure H$_2$/D$_2$ plasmas. A sample placed in the plasma is negatively biased with respect to plasma potential. Upon positive ion impacts on the sample, some negative ions are formed and detected according to their mass and energy by a mass spectrometer placed in front of the sample. The experimental methods developed to study negative ion surface production and obtain negative ion energy and angle distribution functions are first presented. Different diamond materials ranging from nanocrystalline to single crystal layers, either doped with boron or intrinsic, are then investigated and compared with graphite. The negative ion yields obtained are presented as a function of different experimental parameters such as the exposure time, the sample bias which determines the positive ion impact energy and the sample surface temperature. It is concluded from these experiments that the electronic properties of diamond materials, among them the negative electron affinity, seem to be favourable for negative-ion surface production. However, the negative ion yield decreases with the plasma induced defect density.

## Introduction

Negative-ion production on surfaces in low-pressure plasmas rely on two distinct mechanisms. Depending on where the ions are formed, one distinguishes volume production[1,2,3,4,5] associated with dissociative attachment of electrons on molecules and surface production associated with the capture of one or two electrons by neutral atoms or ions impinging on the surface. Depending on the targeted application, either surface or volume production can be the most favourable process. Hydrogen negative-ion sources for fusion[6,7], high energy linear particle accelerators[8,9,10], neutron generation[11], Tandem accelerators and accelerator based mass spectrometry[12,13] all use the principle of enhanced surface production by injection of caesium. Negative-oxygen-ion sources for Secondary Ion Mass Spectrometry (SIMS) operate by volume production. While plasma thrusters for space propulsion[14,15,16] and microelectronics etching plasmas[17,18,19,20,21,22] currently rely on volume production, they may also benefit in future from utilising surface production.

The present work deals with negative-ions for fusion applications in the context of the international projects ITER and DEMO, which aim to demonstrate controlled nuclear fusion for energy production. In tokamaks (nuclear fusion reactors), a plasma composed of deuterium and tritium is magnetically confined and heated to very high temperatures, around $1.5 \cdot 10^8$ K, to overcome the repulsion between deuterium and tritium nuclei and achieve fusion. ITER will be a research device, focusing on the study of 'burning' magnetically-confined fusion plasmas and providing technological solutions for its successor, DEMO. DEMO will be the first nuclear-fusion power-plant prototype producing electrical energy, targeting ~1 GW of electrical power coupled to the grid[23,24]. In the ITER and DEMO devices, the heating of the plasma will mainly be produced by Neutral Beam Injection (NBI). NBIs systems are key components in achieving high fusion energetic-performances. The ITER NBIs are required to inject 1 MeV beams of neutral deuterium atoms (D) into the tokamak, providing plasma heating and current drive. At such high velocities, much larger than typical electron velocities, the probability of electron capture from $D^+$ ions is too low, so that production of D relies on electron detachment from high-intensity $D^-$ beams. $D^-$ negative-ions are produced in a low-pressure plasma source and subsequently extracted and accelerated.

The ITER negative ion source, currently under development at IPP Garching[7,25] in Germany, operates with a high-density, low–pressure inductively coupled plasma. Extracted $D^-$ current density of 200 A/m$^2$, over a large surface of 1.2 m$^2$, with 5-10% uniformity and low co-extracted electron-current (below one electron per negative ion), during long operation period (3600 s) is targeted. To reach such a high $D^-$ negative-ion current, the only up-to-date scientific solution is the use of caesium. Deuterium negative-ions are created at the extraction region by backscattering of positive ions or neutrals on the plasma grid. Deposition of caesium on the grid lowers the material work function and allows for high electron-capture efficiency by incident particles and thus, high negative ion yields. Studies conducted at IPP Garching show that the ITER negative-ion source can reach the required high current densities. However, drawbacks to the use of caesium have been identified. First, the caesium is continuously injected in the source and its consumption is huge, in the range ~5-10 μg/s[26]. Second, caesium diffusion and pollution of the accelerator stage might cause parasitic beams and/or voltage breakdowns and imply a regular and restrictive maintenance in a nuclear environment. Finally, long-term operational stability with caesium appears to be a technological bottleneck requiring a strict, long, difficult and controlled conditioning of the negative-ion source. These issues complicate the operation of the ITER NBI and push towards a strong reduction of caesium consumption or even the development of caesium-free negative-ion sources for DEMO. The aim of the present work is to investigate alternative materials to caesium-coated metals for surface production of high negative-ion yields in low pressure $H_2$ or $D_2$ plasmas.

Surface production of negative-ions in low-pressure caesium-free plasmas is of interest from a fundamental point of view since it is a part of the global plasma dynamics and might influence it strongly. However, few papers [27,28] have been dedicated to this subject, and most of them are related to the process of thin film deposition by magnetron plasma sputtering[29-34], where sputtered negative-ions might influence the properties of the deposited layer. Most of these studies concern O⁻ negative ions. As fast negative ions can potentially damage deposited layers, negative-ion surface production is seen as a drawback in these applications and no attempt has been made to optimise negative-ion yield in these studies. Therefore, in order to find alternative solutions to caesium for fusion applications, there is a need for fundamental studies on negative-ion surface production in low-pressure $H_2$ and $D_2$ caesium-free plasmas. In this context we are studying $H^-/D^-$ surface production on diamond surfaces. The aim is to understand and optimize surface production. The paper is organized in four parts. In the first part the basic principles of negative-ion surface production and possible high negative-ion yield materials materials are briefly presented. In the second part the choice of diamond as material to enhance negative-ion surface production in plasma is justified. The experimental methods are detailed in the third part. The last part is devoted to the study of negative-ion surface production on diamond and summarises both past and recent measurements of diamond's performance. Results for intrinsic or boron-doped microcrystalline, nanocrystalline and single crystal layers will be presented and compared to graphite (Highly Oriented Pyrolitic Graphite, HOPG) used here as a reference material.

## Negative-ion surface production

*Basic mechanisms of negative ion formation at surfaces*

There is extensive literature on negative-ion surface-production in well-controlled beam-experiments at grazing incidence, for a large variety of incident energies and ion particle types and a large variety of surfaces, see for instance references 35-41 and references therein. Negative-ion surface production under quasi-normal incidence has been less studied. From these studies, the fundamental mechanisms of negative-ion surface-production on metallic surfaces have been well established for many years[37,42,43].

Most often beam experiments use positive ions as projectiles, but these are assumed to undergo rapid neutralization on the incoming part of the trajectory leading to the so called memory loss effect. Therefore negative-ion formation can be thought of as the capture of an electron by a neutral projectile irrespective of whether the incoming particle is originally an atom or an ion. Electron transfer from or to a metal surface (see figure 1) is mainly governed by two basic parameters. One is the difference in energy (mismatch) between the affinity level of the negative ion and the Fermi level or the valence band where electrons are to be captured. The second parameter is the coupling between these levels, given by the wave function overlap, which governs the exchange rate. Both ingredients depend on the distance to the surface and the projectile parallel velocity. On approaching the surface the affinity level is **smoothly** downshifted by the image potential, while electron tunnelling transfer rates in both directions between the surface and the projectile affinity-level increases **quasi-exponentially**. At some distance, rates are so large that memory of the initial charge state is lost (the memory loss effect). Close to the surface the projectile equilibrium charge state is a negative ion (figure 1a). However, when leaving the surface, the affinity level rises back and overlaps with empty states in the conduction band (figure 1b), so that the electron will eventually return to the surface, **unless** the rates have sufficiently decreased or the available time is too short, i.e. the velocity too large. This leads to the concept of freezing distance[44] describing the velocity dependent critical distance where electron transfer rates become negligible; the outgoing negative ion fraction reflecting only the local capture and loss rates.

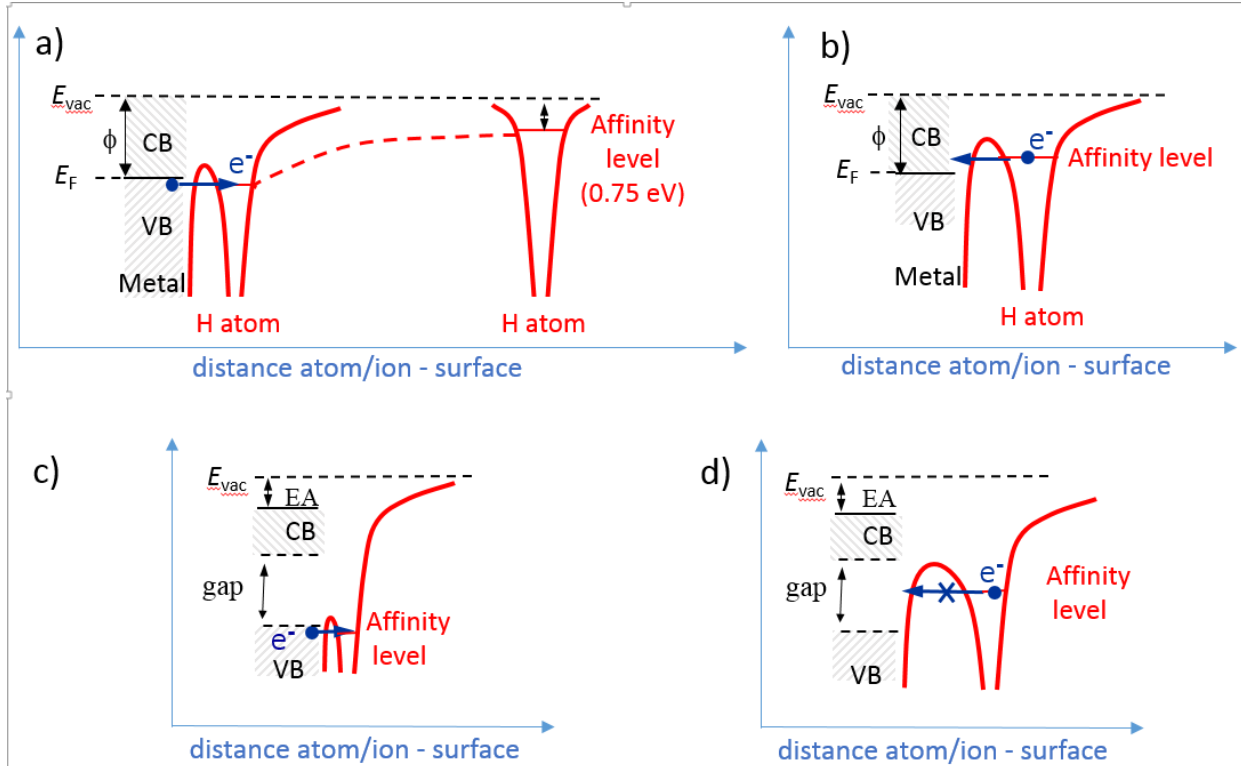

Figure 1: Sketch of the mechanism of electron capture by an incoming hydrogen atom on a metal surface (a, b) and on an insulator (c, d). A) the H atom at large distance is not interacting with the surface; on the approach the affinity level is downshifted; at a short distance, population of the affinity level from occupied states of the material by tunnelling is possible b) as the H- ion is moving away from the surface, the electron can transfer back to empty states of the conduction band c) electron transfer from the insulator valence band to the affinity level takes place at even shorter distance than on metal d) when the H- ion is moving away from the surface, the electron cannot transfer back to the surface because of the band gap. $E_{vac}$ = vacuum level, EF = Fermi level, VB = valence band, CB = conduction band, $\phi$ = work function, EA = electro-affinity

When depositing caesium on the surface, the material work-function is lowered reducing the energy barrier so that the freezing distance is now placed in the region where negative ions dominate. Usual metal work functions are on the order of 5 eV while a thick caesium deposit on the metal will lead to a work function of 2.1 eV (the work function of caesium itself)[45]. The surface work-function can be further reduced in the range of 1.5 eV if only ~half a monolayer of caesium is deposited on the surface[46,47]. However, due to the complexity of a real negative-ion source, there is little chance to obtain such fine control of the caesium coverage[48,49,50]. More interestingly here, if the work-function is low enough, the freezing distance will remain in the favourable region even for low velocity projectiles allowing atoms with ~eV energies to contribute to the negative ion production. This explains the success of the giant negative-ion source for ITER where the atomic flux, which largely exceeds the ionic flux, is able to contribute to negative-ion surface production.

For insulators or semi-conductors, electrons are localized and more deeply bound, resulting in much lower electron capture rates, so that even full neutralization of positive ions is not granted. Still, a general mechanism explaining significant negative ion formation has been identified recently[51]. In addition to the image potential effect, the downshift of the affinity level is further amplified by the Coulomb interaction between the negative ion and the localized hole (figure 1c). Furthermore, on the outgoing part of the trajectory, the electron loss back to the surface can be reduced or suppressed because no empty states are present in the band gap to recapture the electrons (figure 1d). Such models have explained that on LiF(100) surfaces, known to have one of the deepest valence band, yields of 10% $H^-$ have been observed at grazing incidence[52,53], more than one order of magnitude larger than from an Al surface[54].

Aside from this, there are few examples on insulators where negative ion formation occurring directly from positive ions (i.e. simultaneous capture of two electrons), was found to be more likely than single electron capture from neutral atoms[55, 56]. This has been explained with the above quasi molecular model including coulombic attraction and by the fact that the endothermic energy requirement associated to the neutralization of the positive ion can be offset by the exothermic release of energy in populating the affinity level provided that they take place simultaneously. This mechanism is potentially interesting for negative-ion surface production but will not be discussed further in this paper.

*Caesium and its alternatives*

The story of plasma-based negative-ion sources started in 1971 with the discovery of the caesium effect by the Russian scientists V Dudnikov and Y Belchenko[57] (see also review papers 58, 59). Up to 1989, two kinds of negative-ion sources were developed, namely volume sources and surface-plasma sources (SPS). In the former, negative ions are created by attachment of electrons to vibrationnaly excited molecules within a magnetized part of the plasma close to the extraction grid. In the latter, negative ions are created on a negatively-biased cathode facing an extractor grid. The cathode is usually made of low work-function materials and it has been shown that barium can be as efficient as caesium[60,61,62] since the optimum caesium coverage of half a mono-layer cannot be realistically obtained in a real negative-ion source. Furthermore, barium does not present major conditioning issues and its surface can be prepared simply by argon sputtering[63]. In 1989, Leung[64] demonstrated that the injection of caesium vapour in volume sources largely increased the extracted negative-ion current. This new type of source was called a hybrid-source. It took many years of research on Cs-seeded sources before it was understood that the efficiency of the hybrid source is due to a surface effect, mainly arising from the creation of negative-ions on the plasma grid, very close to the extractor[65]. The observed effect is *a priori* not inherent to caesium and could probably be obtained by using other material for the plasma grid. However, for many years studies have focused strongly on caesium and alternative solutions have not been investigated deeply. Therefore, revisiting the efficiency of low-work function materials, other than caesium, in the light of modern negative-ion source developments would be interesting. Such studies have recently started at IPP Garching[66]. Another approach to reducing the drawbacks of caesium would be to limit its injection in the source. Some recent studies[67,68] have shown than caesium-implanted molybdenum can produce high negative-ion yields and could be useful for fusion. Several arguments favour this solution: i) implanted caesium would act as a reservoir for the surface; ii) less than one monolayer of caesium is required to optimize surface production, and this can be achieved by caesium implantation; iii) there is a high chance that the caesium coverage of the extraction grid in an actual negative-ion source is not large even when injecting massive amounts of caesium, since the grid temperature is quite high (150-200 °C[50,7]); iv) the grid could be re-implanted *in-situ* when the caesium reservoir inside the material is depleted. However, further studies are required to check the capability of this technique and verify that large enough caesium surface concentration (few tens of percent) can be reached together with low contamination.

As explained before, insulating materials, in particular large band gap materials, could be interesting for enhancing negative-ion surface production in plasmas. In the ITER negative-ion source, the plasma grid on which extracted negative-ions are formed is biased between the floating and the plasma potential[69]. Therefore using an insulating material at floating potential for the plasma grid would not be *a priori* an issue. An insulating material deposited onto a conductor grid could also be an alternative. Among insulators, we have selected diamond for many reasons. First of all carbon materials appear to be among the best candidates as negative-ion high yield material in caesium free plasmas. A negative-ion yield of 10% on graphite (HOPG, Highly Oriented Pyrolytic Graphite) has been obtained at ISMO[70], and slightly lower yields have been reported on diamond-like carbon (DLC)[71] (yield is defined as the ratio between the negative-ion flux leaving the surface and the positive ion flux impinging on the surface). DLC has even been chosen as a converter material for the low-energy heliospheric neutral atom sensor (neutral to negative-ion converter) installed on the IBEX satellite launched in 2008. Furthermore, in reference 72, the yield measured from diamond was twice that measured from graphite, showing that diamond is a promising negative-ion high yield material. Secondly, we have shown a 5-fold increase in the negative-ion yield from diamond compared to graphite when the diamond surface is heated to ~ 400°C under plasma exposure[73]. Thirdly, diamond can be produced in several forms such as single crystals which usually have small areas of up to 1 cm$^2$ and thicknesses of a few mm [74], polycrystalline films with thickness of 1 to 100 µm, with surface areas of up to 5000 cm$^2$ [74,75] and nanocrystalline films with grain size down to 5 nm[76,77,78]. Tuning of electronic properties can be obtained by deposition of intrinsic, lightly or highly doped layers using boron (p-doped), nitrogen, or phosphorous (n-doped) dopants. In addition, diamond's surface properties depend on the crystallographic orientation of the exposed surface as well as on the termination of the dangling bonds which can be oxygen or hydrogen terminated. Finally, and most interestingly, hydrogenated diamond layers may exhibit negative electron affinity (i.e. the minimum of the conduction band can be above the vacuum level)[79,80] so that any electron in the conduction band is free to leave the surface. Here, the anticipated advantage would be that the limited bandgap implies that the valence band is closer to the vacuum level, favouring electron capture, while still limiting the electron loss back to the surface. Also, if some electrons are promoted into the conduction band by the UV radiations of the plasma, they could be easily captured. The negative-electron affinity of diamond probably explains the excellent field emission capabilities[81] of diamond as well as the observation of very high secondary electron emission yields; up to 80 emitted electrons upon single electron bombardment on a (100) single-crystal with hydrogenated or caesiated surface[82]. Due to the ability of diamond and hydrogenated-diamond to emit a high flux of electrons, diamond is expected to be efficient for producing negative-ions. This last hypothesis is empirical since the basic mechanisms of surface ionisation are different from the mechanism involved in field or secondary electron emission. Nevertheless, a good correlation between *ion-induced* secondary emission yield and O⁻ negative-ion yields has been observed for several materials in a magnetron plasma[30].

## Experimental method

Surface negative ion production measurements are performed in the diffusion chamber of a plasma reactor. We only briefly describe the experiment here, as details can be found elsewhere[83,84]. The plasma is generated at 1 or 2 Pa, either by capacitive coupling from an external antenna using a 20 W 13.56 MHz generator, or at 1 Pa by an Electron Cyclotron Resonance antenna driven by a 60 W, 2.45 GHz generator. The plasma density in the diffusion chamber, as measured by a Langmuir probe, is $n_e = 2·10^{13}$ m⁻³ and the electron temperature is $T_e = 3.5$ eV, giving an ion flux to the sample of the order of $10^{17}$ m⁻²s⁻¹ in RF mode. In the case of an ECR plasma, $n_e = 2.5·10^{15}$ m⁻³, $T_e = 1.0$ eV and the ion flux to the sample is ~$7·10^{18}$ m⁻²s⁻¹. The sample holder lies in the centre of the diffusion chamber, facing a Hiden EQP 300 mass spectrometer equipped with an energy filter. The sample can be biased negatively by an external DC power supply, and can be heated by a resistive heater embedded inside the sample holder (see figure 2b). The sample temperature is monitored by a thermocouple placed at the backside of the sample. The temperature measured by the thermocouple was previously calibrated versus the target surface temperature. The uncertainty on the temperature measurement is estimated to still be quite high (on the order of ±50 K) due to uncertainty in the thermal contact between the sample and the sample holder, depending on how the sample is clamped. During RF plasma, without external heating, the sample temperature is maintained at room temperature because the impinging ion flux is low enough. During ECR plasmas the sample temperature rises by about 70 K due to the higher ion flux. Pristine materials are used for each series of experiments: non-exposed new diamond layers or freshly cleaved HOPG (Highly Oriented Pyrolitic Graphite) samples.

Figure 2a and b show a simple sketch of the experimental arrangement. The sample is negatively biased with respect to the plasma potential so that Negative-Ions (NI) emitted from the surface are accelerated towards the plasma and self-extracted from the plasma to the mass spectrometer, where they are detected according to their energy. This measurement gives the Negative-Ion Energy Distribution Function (NIEDF). Both modelling methods that will be presented later require knowledge of ion trajectories inside the sheaths in front of the mass spectrometer and in front of the sample as a function of ion emission angle and energy. The experimental arrangement has been designed in order to ensure planar sheaths in front of mass spectrometer and sample (the biased surface is much larger than the sample surface to prevent edge effects) [83]. In this situation, the local electric field in the sheaths can be calculated using Child Langmuir law knowing the electron density and temperature (from Langmuir probe measurements) and the surface bias. Ion trajectories in the sheaths are then simply determined by numerically calculating the (Newton's) equation of motion. There are two major advantages to this experimental arrangement. First, samples can be changed easily thanks to a fast load lock system, which would not be possible if one studied materials deposited on an extraction grid. Second, the physics of the ion extraction is quite simple and can be easily accounted for in order to focus on surface production rather than extraction issues. The main disadvantage is the requirement of a negative bias to get the self-extraction effect. Positive ions bombard the sample and create defects. The pristine material is modified and its surface state has to be characterized afterwards.

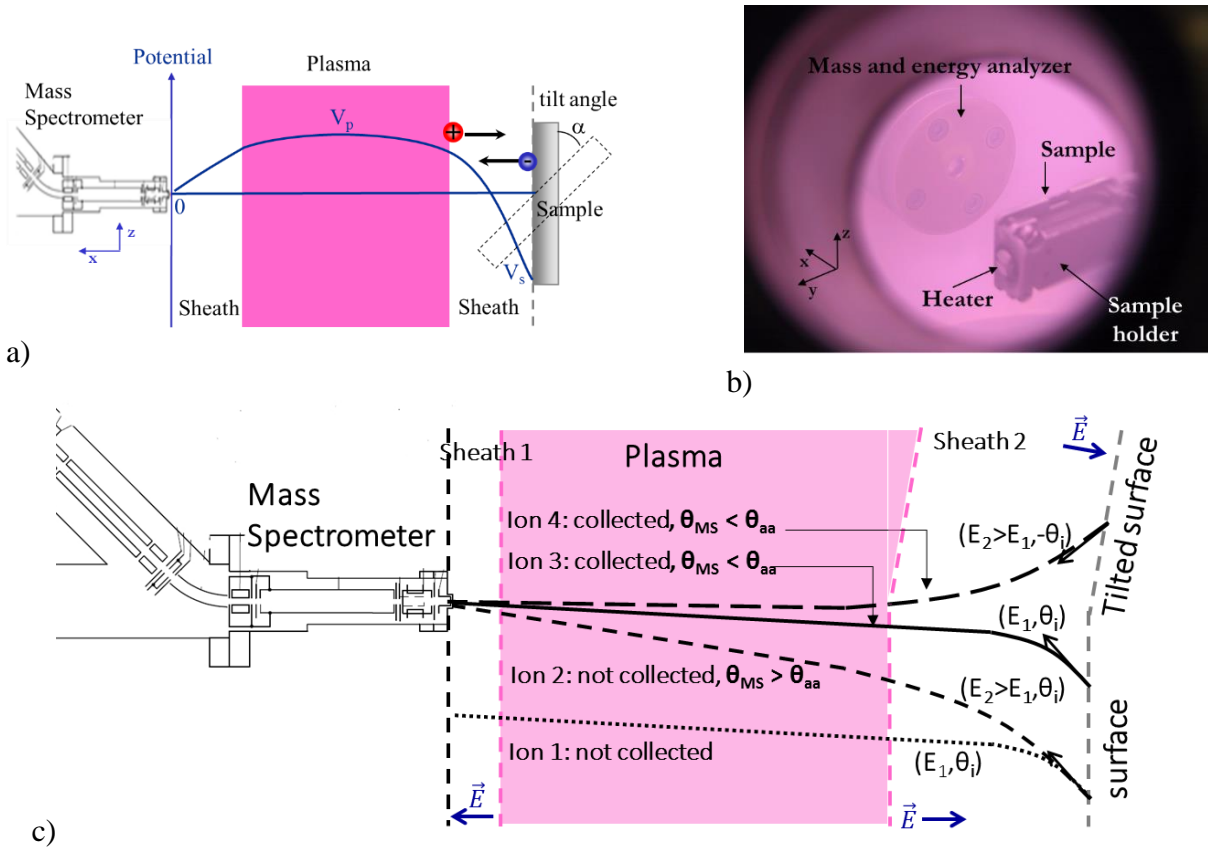

Figure 2: a) Sketch (not to scale) of the experimental arrangement showing the electrical potential profile between the sample holder and the mass spectrometer. b) Picture showing the sample holder facing the mass spectrometer. c) Sketch (not to scale) of trajectories for four negative-ions leaving the surface from three different locations with the same angle and two different energies. Ion 1 is not collected; its energy is low so its trajectory is strongly modified by the electric field in sheath 2, leading it to miss the mass spectrometer (MS) entrance. Ion 3 has got the same energy as ion 1 but its location allows it to reach the MS entrance. Ion 2 reaches the mass spectrometer but at an angle greater than the acceptance angle, so it is not collected. Ion 4 has got the same energy as ion 2 and starts from a radially symmetric location on the sample, but is emitted from a surface tilted with respect to the mass spectrometer sheath. Ion 4 is collected as its arrival angle at the MS entrance is lower than the acceptance angle. For the sake of simplicity, trajectories in sheath 1 are not correctly represented (electric field is much lower in sheath 1).

In the study of negative-ion surface production by beam experiments, most of the information on the surface production mechanisms is concentrated in the scattering differential cross section of the products associated with well-defined momentum of the primary projectile. In plasma the measurement is reduced to the Energy and Angle Distribution Function of the emitted Negative-Ions (NIEADF) which is already difficult to obtain since the negative-ions have to be extracted before being measured. The measured Negative-Ion Energy Distribution Function NIEDF is very different from the NIEDF of emitted ions principally because of the limited acceptance angle of the mass spectrometer that does not allow collecting the full negative-ion flux (see figure 2c). Simulations are needed to take into account this effect. In order to determine the Energy and Angle Distribution Function of the emitted Negative-Ions (NIEADF on the sample surface), we first choose it *a priori*, then calculate ion trajectories inside the sheath based on this choice, and finally produce a NIEDF (at mass spectrometer) restricted to the ions reaching the mass spectrometer within its acceptance angle (see figure 2c). The computed NIEDFs are compared to the experimental ones. The *a priori* choice made for the NIEADF on the surface is validated once a good agreement between the computed and the measured NIEDFs are obtained for different tilt angles. The tilt angle (named hereafter α) is defined on figure 2a and represents the angle between the sample normal and the mass spectrometer axis. When rotating the sample, ions emitted at higher angles and/or higher energies can be collected (see figure 2c and reference 84). This method requires an accurate initial guess of the solution. To generate initial angular and energy distributions (NIEADFs), we used the SRIM[85] software. SRIM is a software package which calculates many features of the transport of ions in matter such as ion stopping and range, ion implantation, ion induced sputtering, etc.. It has been used here to compute the energy and angle distribution functions of hydrogen particles backscattered or sputtered from the surface upon hydrogen positive-ion bombardment[28,86,87,88]. We have assumed that these distributions are those of negative-ions. This is justified by the fact that we have previously shown that under our experimental conditions negative-ions are formed by backscattering of impinging positive-ions and by the sputtering of adsorbed hydrogen atom [28,86]. Ion implantation has been disregarded in the present study as it is not relevant for NIEADF calculations, but one can note that about 80% of the positive ion impacting the surface are implanted and contribute to the hydrogenation of the material under study. More information on SRIM calculations can be found in reference 88. A comparison between computed and experimental NIEDFs is presented in figure 3 for zero tilt angle, considering different hydrogenation of carbon material. Comparisons at different tilt angles are presented in ref 84. The good agreement between calculations and experiments (figure 3 for $\zeta_H = 30\%$, and reference 84) validates the initial choice of the NIEADF. Despite this, we cannot prove the uniqueness of this solution, but the good agreement obtained and the fact that the initial choice of the NIEADF is made upon physical considerations gives confidence in the solution found. As SRIM does not take into account surface ionization, attributing the SRIM energy and angle distribution functions to those of the emitted negative-ions is a strong assumption, implicitly stating that the surface ionization probability has no dependence on the energy and angle of the outgoing particle. In the present context this has been found to be an acceptable assumption for carbon materials[83,84]. The input for the SRIM computations were the positive ion distribution functions, as measured by the mass spectrometer and the surface parameters, namely the hydrogen surface coverage and the hydrogen surface binding energy. Here, we benefited from intensive studies of carbon materials as plasma facing components in tokamaks. SRIM computations for hydrogenated carbon layers have been largely validated. But this cannot, *a priori*, be generalized to any material. We have therefore developed a second method to derive the NIEADF on the sample surface. In this method NIEDFs at mass spectrometer are measured for several tilt angles. When rotating the sample, ions emitted at higher angles and/or higher energies can be measured (see figure 2c and reference 84). By using NIEDFs measured at all tilt angles (figure 4 a), an inverse calculation can be performed to determine the NIEADF without any *a priori* assumption[89]. This computation is a complex inverse problem. It is again impossible to prove the uniqueness of the solution so all precautions have been taken to obtain the most physical and appropriate solution from the inverse calculation. The details are not given here but the reader may refer to reference 89 for more details. The NIEADF on the surface obtained with the second method is presented on figure 4b.

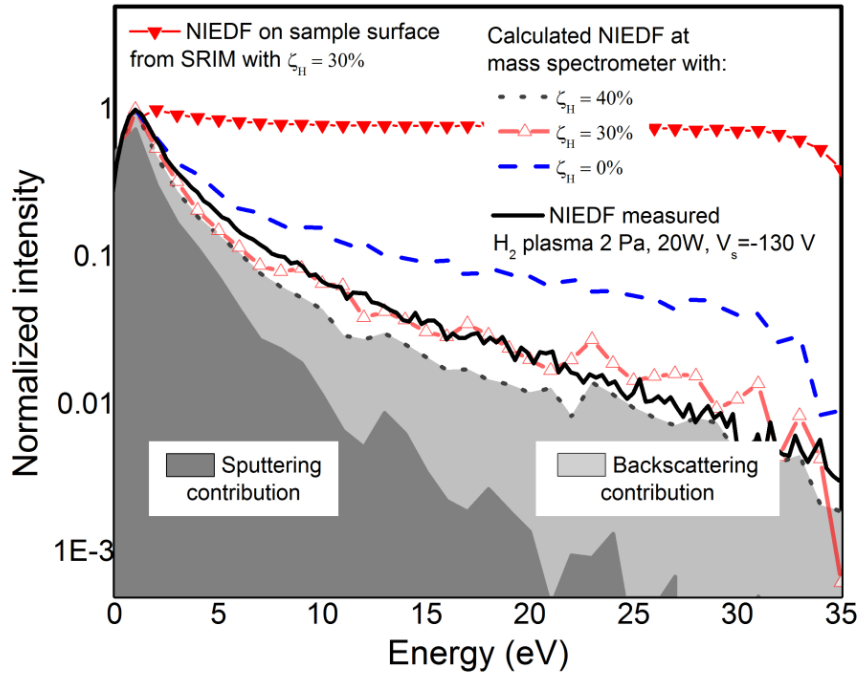

Figure 3: Solid triangles represents the NIEDF on sample surface from SRIM calculations assuming impacts of 50 eV $H^+$ ions (corresponding to 150 eV $H_3^+$ ions dissociated at impact) on hydrogenated carbon material with 30% of hydrogen ($\zeta_H = 30\%$). Dash, Dash-dot and dot lines are calculated NIEDF at the mass spectrometer for different hydrogenation of carbon material ($\zeta_H$ from 0 to 40%). The dash dot distribution has been calculated using the distribution on sample surface shown with solid triangles. The contributions of sputtering and backscattering are shown for $\zeta_H = 40\%$. Also shown is the experimental NIEDF measured with HOPG sample in a $H_2$, 2 Pa, 20 W RF plasma, for a surface bias $V_s$=-130 V. In this situation the plasma is $H_3^+$ dominated and positive ions impact the surface at approximately 150 eV due to the difference between the plasma potential and the surface bias.

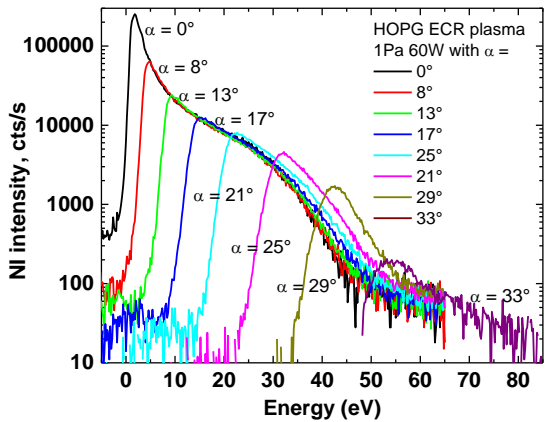 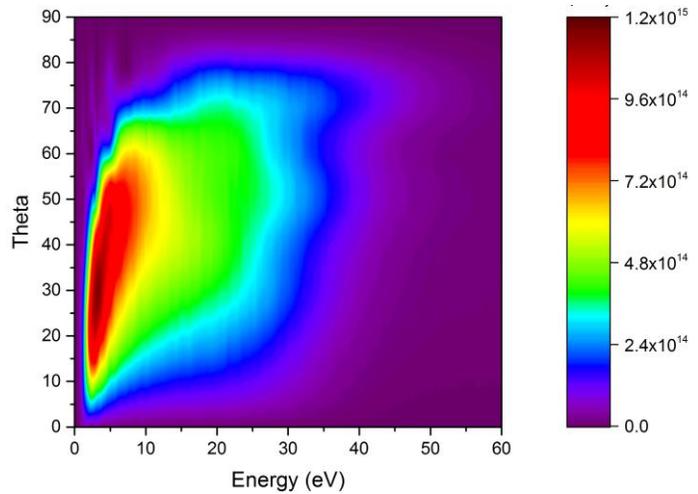

Figure 4a: Experimentally measured mass spectrometer NIEDF for different tilt angles of the HOPG sample: from α = 0° to 35°, with a step of 1°. Measurements were made in an ECR 1 Pa, 60 W plasma.

Figure 4b: NIEADF on the sample surface computed using the NIEDFs measured at mass spectrometer (figure 4a) as input. The colour coded intensity map indicates the number of NI emitted from the surface at an angle θ and with an energy E.

There is an overall good agreement [89] between both methods. The second method can however be generalized to any kind of material and any negative-ion type. The results presented in figure 3 and 4 concern HOPG material but identical results are obtained with diamond layers. Several outputs come from these calculations. Firstly an estimate is obtained of the hydrogen surface coverage, more precisely the coverage of hydrogen participating in the sputtering process leading to negative-ion formation. For both HOPG and diamond layers it was found to be of the order of 30% at room temperature, in RF and ECR plasmas, with a sample bias of -130 V. Furthermore, the calculations showed that under the chosen experimental conditions measured negative ions are not coming from the whole sample surface. Only ions coming from a disc of diameter < 2 mm centred on the 8 mm diameter sample can be collected by the mass spectrometer. Ions coming from outside this area and reaching the mass spectrometer entrance arrive at an angle greater than the acceptance angle and are not collected. This is illustrated on figure 2c. When the sample is tilted, the centre of the collection disc slightly shifts but never reaches the edge of the sample, even at 35° tilt angle[84] (the centre shifts by about 1.5 mm at 30°). The clamp and the sample holder surfaces therefore do not contribute to the total yield of negative ions. The calculations also show that the sputtered negative ions are emitted at lower angle and energy than backscattered ions and hence those are preferentially detected when the sample surface is normal to the mass spectrometer axis (α = 0°). While 95% of ions are emitted by the backscattering process, about 40% of ions being detected when α = 0° have been created by the sputtering process. Moreover, only a few percent of the emitted negative-ions are detected (e.g. 1.6% for a 2 Pa, 20 W RF plasma). Rotating the sample is therefore crucial to collect information representing the total negative-ions distribution. When comparing two materials on the sole basis of the distribution recorded at α = 0°, one must ensure that their angular emissions are identical. This verification has been made for all measurements shown in this paper. The models also demonstrate that the shapes of the measured energy distribution functions are mainly determined by the proportion of sputtered and backscattered negative ions detected as demonstrated on figure 3. When tilting the sample, the main peak, originating from sputtered negative-ions, disappears[84] and negative-ions originating from backscattering mechanism are probed. Finally, the models show that the measured NIEDFs (figure 3 and 4 a) are very different from the distribution functions of the particles emitted by the surface[84]. It is illustrated on figure 3 where NIEDF on sample surface and at mass spectrometer are shown in the case of a hydrogenation of 30% (solid triangles and dash dot line). This difference is due to the low acceptance angle of the MS that limits the collection of NI to only a part of the total NI flux.

## Diamond as a negative-ion high yield material: results and discussion

Figure 5 compares relative negative-ion yields obtained on graphite (HOPG) and diamond materials in $D_2$ RF plasma (2 Pa, 20 W, surface bias $V_s = -130$ V) for different surface temperatures. The NI yield is usually defined as the ratio between the negative-ion flux leaving the surface and the positive ion flux impinging on the surface. Under the present experimental conditions the positive-ion flux is constant. Therefore, the relative NI yield is simply defined as the measured total flux (counts per second) of negative-ions (i.e. the area below the measured NIEDF). It has been checked that the angular emission behaviour of all the layers studied is identical, hence a comparison of yields measured at $\alpha = 0°$ is enough. As it has already been demonstrated in $H_2$ plasma, the negative-ion yield on diamond exhibits a maximum around 400–500°C while the yield on graphite is continuously decreasing (see figure 5, continuous bias curves). Two diamond layers are presented here: i)Micro-Crystalline Boron-Doped Diamond (MCBDD), whose Boron doping is estimated to be $1.5 \times 10^{21}$ cm$^{-3}$ (SEM pictures and information can be found in reference 90); ii)Micro-Crystalline Diamond (MCD), which is very similar to MCBDD but without boron doping (see SEM picture on figure 6 where a strongly multitwinned polycrystalline diamond film is shown). More materials are compared in reference 91. By comparing doped and non-doped microcrystalline diamond layers, it can be observed that boron doping does not seem to influence either the negative-ion yield or the global behaviour of the yield with temperature. One can notice that no data points have been obtained for MCD at temperatures lower than 300°C. The reason is that the un-doped MCD layer turned out to be insulating at lower temperatures when exposed to $H_2$ plasma (2 Pa, 20W RF). As the self-extraction method requires a negative DC bias of the sample, it was not possible to undergo any measurement on MCD from room temperature to 300°C. We therefore conclude that boron doping simplifies the present study by giving a good sample conductivity but does not seem to influence the NI yield.

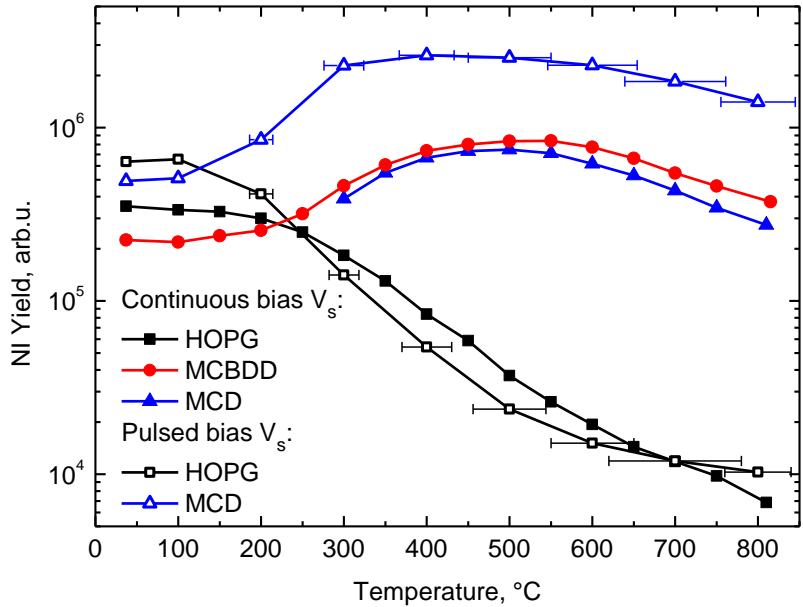

Figure 5: NI yield dependence on the surface temperature for HOPG, MCBDD and MCD for constant bias (solid symbols) and pulsed bias (empty symbols). Plasma parameters: 2.0 Pa of $D_2$ RF plasma, 20 W. Pulsed bias parameters: $T_{pulse}$ = 15 µs, $T_{acq}$ = 10 µs, f = 10 kHz, $V_s$ = –130 V.

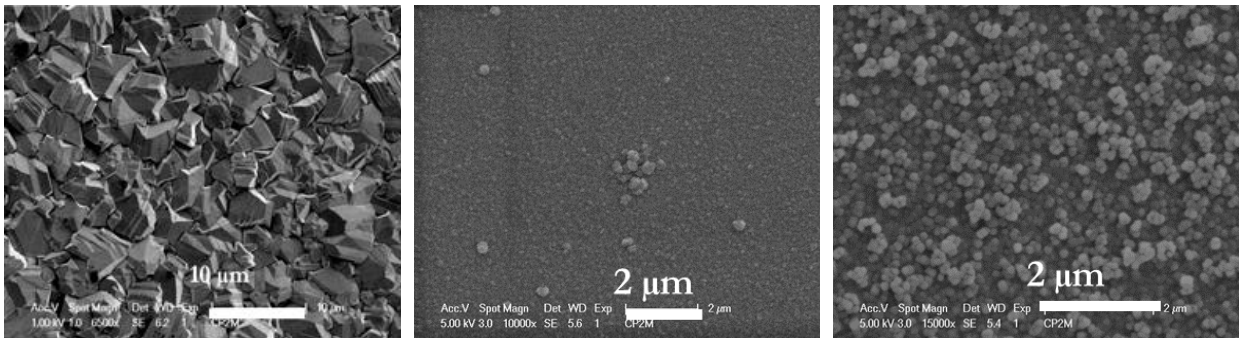

Figure 6: SEM picture of MCD material (micro-crystalline diamond), NCD 1% and NCD 5% deposited on Si at LSPM laboratory. Thickness is from 2 to 10 µm for MCD and around 200 nm for NCD depending on the sample used.

The increase of the negative-ion yield with surface temperature is still under study. However, surface analyses, in particular Raman spectroscopy, have provided insight into the understanding of this behavior[73,90]. At room temperature, the ion bombardment creates some defects on the material and lead to its hydrogenation. The positive ion energy in figure 5 is around 135 eV (plasma potential minus surface bias). As $H_3^+$ is dominating the positive ion flux, the energy per nucleon of positive ions impinging the surface is around 135/3 = 45 eV. The impacts lead to the creation of a hydrogenated carbon layer having a certain hybridization ratio $sp^2/sp^3$, most probably higher for graphite than for diamond. Raman spectra reveal that $sp^2$ phases on the MCBDD surface disappear with the temperature increase. Defects/non diamond phases produced by plasma exposure are annealed and etched away by the plasma at high temperature. Enhanced $sp^2$ phases etching at high temperature has indeed already been observed[92,93]. Therefore when increasing the temperature, the diamond surface is reconstructed and the surface state under plasma exposure is probably closer to the pristine material than it would be at room temperature. At 800°C, the diamond layer almost recovers its original Raman signature. From the modelling presented previously we learn that up to the maximum in negative-ion yield, the hydrogen surface coverage is slightly increasing (from 30 to 35%) while it decreases after the maximum. H-free (100) diamond surface has shown positive electron-affinity contrary to hydrogenated (100) surfaces which presents negative electron-affinity[79]. Therefore we can assume that upon heating from room temperature to 400-500°C under plasma exposure the diamond layer recovers its electronic properties and among them, its negative-electron affinity. When further increasing the temperature, hydrogen atoms desorb and the negative-electron affinity is lost. Concerning graphite, Raman measurements show that its surface is reconstructed with increasing temperature, while modelling demonstrates that hydrogen coverage drops to zero. However, the decrease of the negative-ion yield is much bigger than that expected from the complete suppression of the sputtering mechanism. Therefore, one can conclude that $sp^2$ hybridization is not favourable for negative-ion surface creation under our experimental conditions (i.e. under high energy ion impacts at normal incidence). This is confirmed by the time evolution of the negative-ion yield under $D_2$ plasma exposure presented in figure 7a. Both materials were heated to 500°C under vacuum prior to plasma exposure in order to remove impurity contamination. The samples were then kept under vacuum until the surface temperature returned to room temperature. This precaution prevented any change in negative-ion yield due to impurity cleaning during first few minutes of plasma. Let us note that under the present experimental condition (20 W RF plasma) the positive ion flux is low and the surface remains at room temperature under plasma exposure. The evolution observed is thus not due to change of surface temperature under plasma exposure. Once the plasma is started, some defects are created on the surface ($sp^3$ defects for HOPG, $sp^2$ defects for Diamond) and hydrogen is implanted. One can observe that the negative-ion yield is increasing for HOPG (the creation of $sp^3$ defects favours negative-ion creation), while it is strongly decreasing for MCBDD (the creation of $sp^2$ defects is not favourable). Figure 7b shows a similar measurement in deuterium plasma. The empty symbols represent the negative-ion yield from the first series of measurement on a virgin MCBDD sample. One can observe a major NI yield decrease during the first 5 minutes (see black empty square symbols) probably connected to the degradation of the sample surface. Once the surface state has stabilised, the negative-ion yield stays constant. The sample was then heated to 400°C under plasma exposure with a bias at $V_s$ = -130 V. A stable temperature of 400°C was reached after about one minute. As can be seen from the red empty circle symbols, the negative-ion yield increases significantly during the first 5 minutes the sample is held at 400°C. The negative-ion surface production becomes more efficient and the yield rises to slightly below the initial level. This can be interpreted as the surface state partially recovering from the defects induced during the plasma exposure at room temperature. In order to check if heating to 400°C reconstructs the surface completely, a second series of measurements on a second MCBDD sample (MCBDD 2) was carried out and results are displayed as solid symbols in figure 7(b). The sample was first heated to 400°C before immersion in the plasma and a decrease of the negative-ion yield was still observed. However, the yield went down by a factor of 2 instead of the factor of 3 seen previously, and the surface state changed more gradually, taking ~ 20 min to stabilise. This demonstrates that heating of MCBDD hinders the creation of defects under plasma exposure, keeping the surface in a state which is more favourable for negative-ion production. The red solid circle symbols show the continuation of the experiment with the MCBDD 2 sample after letting it cool down overnight in vacuum. One can notice that the surface that was previously exposed at high temperature (black solid square symbols) initially presents the same negative-ion yield at room temperature as an unexposed sample (red solid circle symbols), showing that the surface state after heating to 400°C is close to an undisturbed surface state. For the second half of the experiment, illustrated with red solid circles and green solid triangle symbols, the MCBDD 2 sample sample undergoes the same exposure and heating cycles as the MCBDD 1, showing similar behaviour. The 15% difference in the negative-ion yields between MCBDD 1 and MCBDD 2 samples under the same conditions is thought to be due to the experimental uncertainty in the surface temperature and alignment of the normal to the sample surface with respect to the mass spectrometry axis.

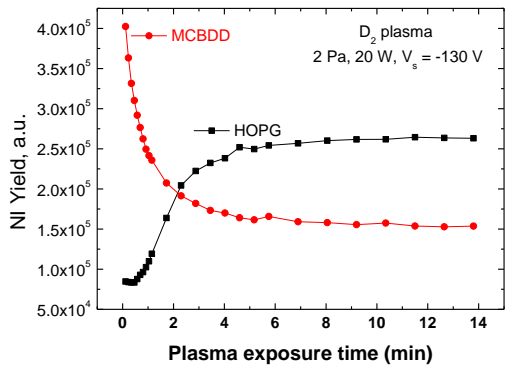

Figure 7a: Time evolution of the negative-ion yield for HOPG and MCBDD materials from the onset of the plasma. Each sample was heated under vacuum to 500°C for 5 minutes to release impurities before being returned to room temperature for the start of experiments.. Each sample's temperature was 300 K (room temperature) while it was biased at $V_s = -130$ V and exposed to a $D_2$, 2 Pa, 20W plasma.

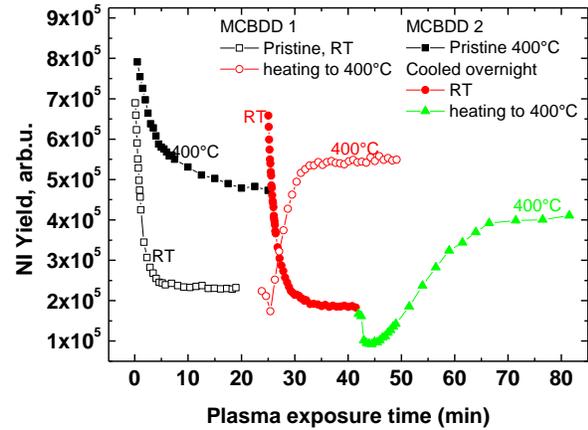

Figure 7b: Time evolution of the negative-ion yield from MCBDD under plasma exposure at different surface temperatures. Each sample was heated under vacuum to 500°C for 5 minutes to release impurities before being returned to room temperature for the start of experiments. Empty symbols correspond to the MCBDD 1 sample and solid symbols to MCBDD 2. The colour of symbols indicate the chronological order of the experiments: black, red, green. The sample surface temperature during plasma exposure is indicated next to the curves with a corresponding colour. The samples were biased at $V_s = -130$ V in a $D_2$, 2 Pa, 20W plasma.

All these results show that it would be interesting to work with less defective diamond layers. This can be achieved if the positive-ion energy is strongly reduced. Such a situation is relevant for fusion since in the ITER negative-ion source the plasma grid is biased only a few volts below the plasma potential. However, under our experimental conditions, the bias cannot be reduced too much since self-extraction of negative-ions is required. We found that working with a bias of -20 V is possible and that negative ions are still efficiently extracted from the plasma. As the plasma potential is higher with this low bias, the impinging positive ion energy becomes 36 eV. $H_3^+$ ions dominate the ion flux giving an impact energy of ~12 eV/nucleon. Let us note first that this energy is still large enough to create some defects. Indeed, we have conducted NI yield time evolution measurements at $V_s$ = -10 V (corresponding to 9 eV/nucleon) and still observed a yield decrease of the same magnitude as the one displayed on figure 7(a). The mass spectrometer entrance is usually set at 0 V which is a requirement for the NIEDF modelling[83]. However, the mass spectrometer entrance can be biased to positive values when modelling is not needed. We biased it at + 10V and kept the sample surface at ground potential leading to an impact energy of ~5 eV/nucleon. In this case we observed no degradation, with in fact a slight increase in the NI yield with time. However, the signal was too low for a complete study, and a bias voltage of -20 V was chosen as representative of a situation where the positive ion flux does not induce too many defects. Figure 8 presents the NI yields obtained from different diamond layers as a function of surface temperature. In addition to the previous materials HOPG and MCBDD, two different Nano Crystalline Diamond layers have been used here, as well as a (100) Single Crystal Boron Doped Diamond (SCBDD) sample. This 20 µm thick, 9 mm$^2$ area SCBDD crystal was grown on a SUMITOMO HPHT diamond substrate. It was highly boron doped ($10^{21}$ Boron atom/cm$^3$). NCD 1% and NCD 5% refer to the percentages of $CO_2$ used in the gas mixture during the deposition process. The two different layers have very similar Raman signatures but differ in their grain sizes: a few tens of nm and around 200 nm respectively (see SEM pictures on figure 6).

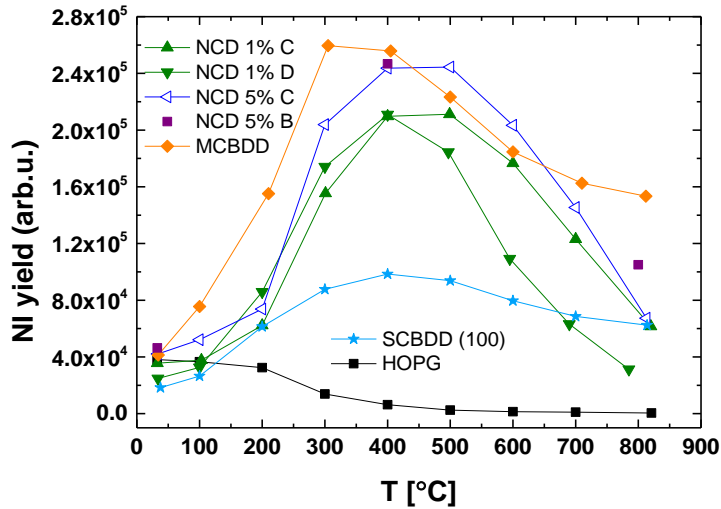

Figure 8: NI yield dependence on the surface temperature for HOPG, MCBDD, NCD 1%, NCD 5% and SCD (100). All diamond materials have been synthesized at LSPM laboratory (CNRS, Paris 13 University). Labels B, C and D refer to different samples of the same material. $D_2$ RF plasma, 20W, 2 Pa, $V_s$ = - 20V.

First of all, figure 8 demonstrates the same global behaviour of negative-ion yield versus surface temperature that is observed at higher positive ion energy. The yield is decreasing for HOPG while it is increasing for all diamond layers up to a maximum around 400 - 500°C after which it decreases. Again, we can infer that the increase in surface temperature favours surface reconstruction and the recovery of pristine diamond electronic properties. Second, one can observe some dispersion in the NCD measurements. NCD 1% samples C and D gave different negative-ion yields at high temperature, despite both coming from the same original larger sample that was cut into smaller pieces. We have observed that NCD layers can handle several heating cycles without degrading their yields. Let us note that NCD layers are deposited at much lower temperatures than MCD samples: less than 400°C compared to 800°C for MCD layers[94]. Finally, we can observe a similar negative-ion yield behaviour for (100) Boron Doped Single Crystal Diamond as for micro- and nanocrystalline layers. The yield is much lower but this is probably due to a reduced size of 3 mm by 3 mm for the (100) SCBDD sample, which makes its installation on the sample holder and its alignment with the mass spectrometer somewhat complicated. It might even be possible that part of the signal measured for SCBDD comes from the molybdenum sample holder instead of the diamond layer explaining the low signal recorded.

From all the measurements presented here it is not possible to observe a clear influence of the crystalline structure of the diamond layer on the negative-ion yield. Most often the MCBDD or MCD samples produce a slightly higher yield than other materials but overall, a similar behaviour is observed. NCD layers are interesting since they can be easily deposited on large surfaces at relatively low-temperatures and could possibly be regenerated *in-situ* inside a negative-ion source. However, they are expected to exhibit lower resistance to plasma exposure than microcrystalline layers because of their higher sp2 phase content[95,96]. The main purpose of the present study is not to choose between MCD or NCD, but rather to demonstrate if the electronic properties of diamond, in particular negative-electron affinity, influence negative-ion yields, in order to optimise the negative-ion yield by tuning and designing the best material. We have shown that the less defective the diamond surface is, the higher the negative-ion yield is. Therefore, it seems that diamond's electronic properties are relevant for negative-ion surface production, a suggestion supported by the following paragraphs.

As seen before, the MCD material cannot be DC biased when exposed to plasma at low temperature since the layer is insulating. In order to study insulating materials, we have developed a pulsed-DC bias scheme similar to those described in [97] and [98] for the sputtering of insulating films or for the measurement of the positive ion fluxes to plasma chamber walls. The detailed method will be described in a forthcoming paper. The principle is to apply to the insulating sample a short negative DC bias. As the sample is initially not charged, the applied bias appears on the surface. Positive ions from the plasma are attracted to the sample surface decreasing the surface bias. The rate of decrease of the surface bias in Volts/seconds is given by the ratio between the ion saturation current at the sample and the sample capacitance. Under our experimental conditions the ion saturation current never exceeds 100 µA/cm$^2$ and the diamond layer capacitance is of the order of 1 nF corresponding to a 1 cm$^2$ sample of 5 µm thickness. This gives a surface bias decrease rate on the order of 0.1 V/µs. If the measurement is fast enough, the surface bias is almost constant during the measurement. The time resolution of the mass spectrometer being 2 µs, the measurement can be performed at almost constant bias. This allows insulating materials to be studied with the tools developed for conductive materials. Finally, when the bias is switched off, the electrons come to the surface to discharge it before the next pulse starts.

In order to compare the pulsed bias method with continuous bias conditions it was first applied to HOPG at pulsed bias conditions given by: $V_s$ =-130 V, 2 Pa, $D_2$, 20 W, 15 µs pulse, 10 kHz repetition frequency. The NI acquisition time has been set to 10 µs giving a sample surface bias variation of about 1 V during measurement (on MCD), which is low enough to avoid perturbation of the NIEDF. The 10 µs acquisition period starts after the bias has been applied for 5 µs in order to make sure the positive ion flux has stabilised (stabilisation should occur after about 2 µs[99]). Therefore, during pulsed measurements, the positive ion flux is assumed to be equal to that during continuous bias measurements. Interestingly, HOPG has demonstrated higher yields at room temperature under pulsed-DC bias than under DC bias conditions (figure 5). The analysis made with the model presented previously showed that under pulsed bias conditions the HOPG surface is more hydrogenated, leading to a significant amount of NIs created by sputtering. A pulsed bias temperature scan has been performed for the HOPG and MCD materials. The results are presented in figure 5. It can be seen that measurements on MCD have been successfully extended to low temperatures. The global behaviour with temperature is similar to the DC bias mode: the yield from HOPG decreases with increasing temperature and is similar in magnitude to before, while the diamond material presents a maximum in yield around 400°C, as before. However, the yield from MCD is increased by a factor 2 to 5 compared to DC bias mode. If compared to HOPG at room temperature, the yield from MCD is almost one order of magnitude higher in pulsed mode at 400°C.

In the pulsed-DC bias scheme, the positive-ion bombardment only occurs during a short period of time. It is suggested that the increase of NI yield under pulsed bias conditions is the consequence of a less degraded surface with properties closer to a pristine diamond one. This is consistent with the previous studies demonstrating that surface defects tend to decrease the negative-ion yield. Also, as the surface is not conductive, some electrons coming from the plasma when the bias is off, might be trapped in defects in the band gap and might contribute to negative-ion surface production when the bias is ON. The identification of the exact electron-capture mechanism by incoming hydrogen ions still requires further investigation; it will be essential to optimize surface production from diamond or any other material. It should be noted that the applicability of pulsed bias mode is limited by the duty cycle which here is 15%. Nonetheless results show that diamond's electronic properties are promising for negative-ion surface production and that there is still room for optimization of negative-ion yields from these diamond layers.

## Conclusion

The present papers deals with $H^-/D^-$ negative-ion surface production in low-pressure $H_2/D_2$ plasmas. Different diamond materials ranging from nanocrystalline to single crystal layers, either doped with boron or intrinsic have been investigated and compared with HOPG. Measurements were performed in DC or pulsed bias conditions as a function of surface temperature. Boron doping eliminates the conductivity problem and was found to have no influence on negative-ion surface production. As for the crystalline structure, we have not been able to observe a clear effect. However we emphasise that the creation of defects on the diamond surface by positive ion bombardment reduces the negative-ion surface production. We concluded that, compared with HOPG, the electronic properties of diamond, principally its negative electron-affinity, are probably responsible for the observed increase of the negative-ion surface production.

## Acknowledgments

This work was carried out within the framework of the French Research Federation for Fusion Studies (FR-FCM) and the EUROfusion Consortium and has received funding from the Euratom research and training programme 2014-2018 under grant agreement No 633053. The views and opinions expressed herein do not necessarily reflect those of the European Commission. Financial support was received from the French Research Agency within the framework of the projects ITER-NIS 08-BLAN-0047, and H INDEX TRIPLED 13-BS09-0017. PACA county is gratefully acknowledged for its financial support through project "PACAGING 2012_10357". CGI (Commissariat à l'Investissement d'Avenir) is gratefully acknowledged for its financial support through Labex SEAM (Science and Engineering for Advanced Materials and devices) ANR 11 LABX 086, IDEX 05 02